# Recommended high performance telescope system design for the TianQin project

ZICHAO FAN[1], LUJIA ZHAO[1], JIANGUO PENG[1], HUIRU JI[2], ZHENGBO ZHU[2], SHILI WEI[2], YAN MO[1], HANYUAN CHEN[2], AND DONGLIN MA[1,2,] *

[1]*MOE Key Laboratory of Fundamental Physical Quantities Measurement & Hubei Key Laboratory of Gravitation and Quantum Physics, PGMF and School of Physics, Huazhong University of Science and Technology, Wuhan 430074, China*
[2]*School of Optical and Electronic Information and Wuhan National Laboratory for Optoelectronics, Huazhong University of Science and Technology, Wuhan 430074, China*
*\*madonglin@hust.edu.cn*

**Abstract:** China is planning to construct a new space-borne gravitational-wave (GW) observatory, the TianQin project, in which the spaceborne telescope is an important component in laser interferometry. The telescope is aimed to transmit laser beams between the spacecrafts for the measurement of the displacements between proof-masses in long arms. The telescope should have ultra-small wavefront deviation to minimize noise caused by pointing error, ultra-stable structure to minimize optical path noise caused by temperature jitter, ultra-high stray light suppression ability to eliminate background noise. In this paper, we realize a telescope system design with ultra-stable structure as well as ultra-low wavefront distortion for the space-based GW detection mission. The design requirements demand extreme control of high image quality and extraordinary stray light suppression ability. Based on the primary aberration theory, the initial structure design of the mentioned four-mirror optical system is explored. After optimization, the maximum RMS wavefront error is less than $\lambda/300$ over the full field of view (FOV), which meets the noise budget on the telescope design. The stray light noise caused by the back reflection of the telescope is also analyzed. The noise at the position of optical bench is less than $10^{-10}$ of the transmitted power, satisfying the requirements of space gravitational-wave detection. We believe that our design can be a good candidate for TianQin project, and can also be a good guide for the space telescope design in any other similar science project.

## 1. Introduction

Gravitational-wave (GW) is the product of the perturbation of the gravitational field, which was originally predicted by Einstein soon after proposing his theory of general relativity. In 1974, the existence of GW was indirectly demonstrated by the discovery of an orbiting binary system of two neutron stars roughly the mass of the sun. In 2015, the GW signal GW150914 was detected by the ground-based laser interferometer GW observatory (LIGO-VIRGO), this discovery directly confirmed Einstein's prediction [1].

Ground-based interferometric GW observatories, such as LIGO, Virgo, and GEO-600 [2-4], have arm lengths on the order of kilometers and suffer from many sources of terrestrial noise, making it difficult to detect low-frequency GW signals below 1 Hz. In order to break through the limitation of the ground-based system, a space GW detection scheme with the length of the interference arm reaching $10^5$ km was proposed. The Laser Interferometer Space Antenna (LISA), which is a cooperative mission leading by European Space Agency (ESA) and National Aeronautics and Space Administration (NASA). It was originally proposed in the 1990s and is currently scheduled to launch in 2034 [5]. The LISA mission consists of a constellation of three identical spacecrafts that follow the earth around the sun in a flat equilateral triangle. So far, the three spacecrafts have been chosen to be separated from each other by distances of about 2.5 million kilometers. There are two proof-masses in each spacecraft, so it is also equipped with two telescopes for laser transmission. The spacecraft sends laser beams to and receives laser beams from its matching spacecraft via spaceborne telescopes [6-7].

Recently, Luo et al. proposed a new space-borne GW observatory, the Tianqin project. The TianQin project aims to study a plethora of GW sources in the millihertz (mHz) range (i.e., 0.1~100 mHz) such as ultra-compact galactic binaries and coalescing massive black holes [8]. The space-based GW observatory consists of an equilateral triangle of three spacecrafts which orbit around the earth. The geocentric orbits have certain advantages in engineering such as satellite deployment and easy communication with the ground. But for spaceborne telescopes, the stability (optical path stability and stray light stability) and wavefront quality must be further improved compared with the orbit around the sun due to the thermal stability [9]. The accuracy of the GW measurement is affected directly by these disturbances, as it will introduce the phase noise to the interferometric beat signal.

Plenty of researches on the telescopes have been carried out by LISA [6-7,10]. The telescope is designed based on the Cassegrain telescope, but the on-axis or off-axis design has not been decided. The initial LISA design used an on-axis Ritchey-Chretien (RC) telescope, where the light is reflected by the secondary mirror (SM) and passes through a matching lens directly into the optical bench. Compared with off-axis options, these on-axis designs have several major advantages in terms of thermal stability, volume, mass, and cost. Unfortunately, stray light of the optical system, especially the back-scattering from the SM, made LISA later turn to an off-axis design. Even though some efforts are proposed to suppress the backscattering, it is obvious that the off-axis design is more reliable [11]. Therefore, the off-axis design was the preferred option, and the off-axis design was also adopted in the test prototype for LISA [12]. The current LISA telescope baseline design consists of a two-mirror telescope with a two-lens ocular, with a six-mirror design as an alternative [13].

Telescopes for GW detection are aimed to deliver laser light efficiently from one spacecraft to another for supporting precision metrology beyond the usual requirements for good image formation. As the wavefront of the laser beams will diffract to large angles in the far-field, aberrations from the designed telescope directly affect the optical intensity at the location of the spacecraft far away. Besides, the wavefront error from the telescope will be directly imparted onto the received laser beam, which will lead to a reduction of the signal-to-noise. This also explains why we choose the wavefront error as the performance metric rather than the circled energy to evaluate the telescope during the design process. Obviously, the geocentric orbit in TianQin project brings a more stringent challenge to telescope design, requiring a more perfect wavefront structure as well as a super lower stray light level.

In this work, we describe the performance requirements and many important optical characteristics of a catoptric telescope design aimed at the interferometric study of GWs. As the wavefront quality is of crucial importance for the telescope, a coaxial mirror design was developed based on Seidel aberration theory. Then the occlusion is eliminated by making both the field of view and the aperture off-axis, and the exit pupil direction of the system can be controlled by rotating the eyepiece. The exit pupil of the whole optical system must be located between the quaternary mirror (QM) and the entrance of the optical bench since the aperture at the exit pupil plays an important role in suppressing stray light. A progressive optimization is implemented to gradually decrease the wavefront error over the full FOV. The final design result is obtained through the comparison of two optimized routes. In addition, we performed a sensitivity analysis on the telescope to ensure that the telescope could be manufactured correctly within reasonable tolerances. Finally, we conduct the stray light analysis results in the commercially optical software ASAP, demonstrating the stray light performance of the designed telescope.

## 2. Design Specifications

The Tianqin telescope plays a significant role in the interferometry, thus stringent requirements of high dimensional stability and low back-scattered light are directly from the science requirements in Tianqin project. The key specifications of the telescope are listed in Table 1. The requirement for wavefront quality is unusual and challenging for a space-based GW telescope. The Tianqin constellation orbits the earth, which puts forward more stringent requirements to pointing accuracy. As a result, the sub-allocation of the RMS wavefront error attributed to the telescope design is reduced to λ/300, corresponding to roughly 3.55nm. Another vital requirement is the power of the back-scattering light, which must be kept below $10^{-10}$ of transmit laser power to suppress the phase noise. For the purpose of mitigating the consequences of stray light, an off-axis telescope design is preferred as the back-reflection from the SM of an on-axis design transmits directly back to the detector on the optical bench.

Table 1. Specifications of the Tianqin telescope

| Parameter | Specification |
|---|---|
| Wavelength | 1064nm |
| Wavefront quality (design residual) | ≤ λ/300 RMS@1064nm |
| FOV | ±200μrad |
| Entrance pupil diameter | 220mm |
| Afocal magnification | 40X |
| Stop location | Primary mirror |
| Scattered light power | < $10^{-10}$ of laser power |
| Optical path length stability | 1 pm/Hz$^{1/2}$ @0.1mHz-1Hz |

## 3. Initial structure design

*3.1 Design of primary mirror and secondary mirror*

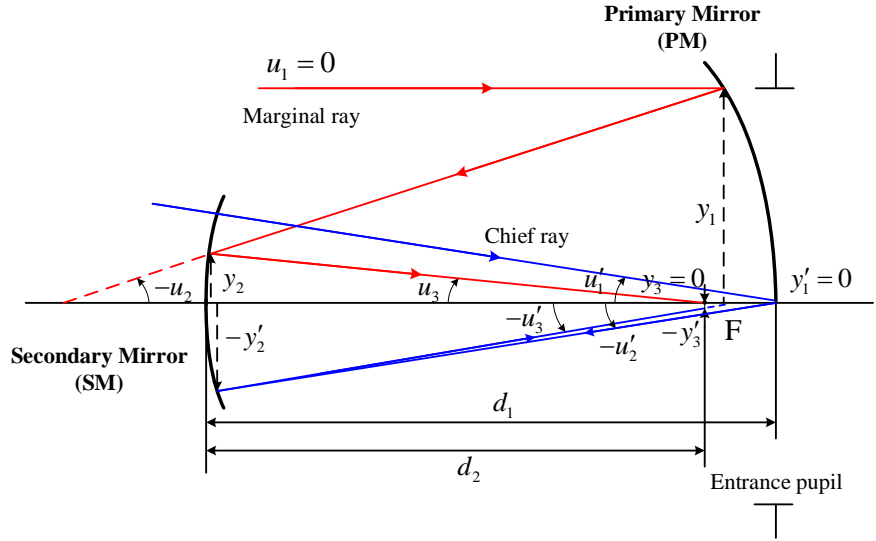

Fig. 1. Characteristics ray coordinates in the initial coaxial system of PM and SM

The Tianqin telescope is designed based on the off-axis four mirrors configuration, which can be temporarily considered as two separate telescopes in the preliminary design phase. Figure 1 shows a ray tracing of an on-axis Cassegrain design suitable for the proposed the TianQin mission. The primary mirror (PM) is nearly parabolic and the SM is aspherical. The even aspherical surface is defined by the following equation:

$$z(x,y) = \frac{c(x^2+y^2)}{1+\sqrt{1-(1+\kappa)c^2(x^2+y^2)}} + \sum c_k(x^2+y^2)^{k+1}, \tag{1}$$

where $\kappa$ is the conic constant, $c$ represents the curvature, $c_k$ stands for the coefficients of the higher-order term. Here is a rotationally symmetric system, $r^2 = x^2 + y^2$. If the primary wave aberrations are described in terms of the Seidel coefficients, the well-known equation can be obtained:

$$W(H,\rho,\theta) = \frac{1}{8}S_I\rho^4 + \frac{1}{2}S_{II}H\rho^3\cos\varphi + \frac{1}{2}S_{III}H^2\rho^2\cos^2\varphi$$
$$+ \frac{1}{4}(S_{III}+S_{IV})H^2\rho^2 + \frac{1}{2}S_V H^3\rho\cos\varphi. \tag{2}$$

The Seidel coefficients $S_{Isph} - S_{Vsph}$ of a given optical system can be calculated by the follow equations [14]:

$$\begin{cases} S_{Isph} = -\sum A^2 \cdot y \cdot \Delta(\frac{u}{n}) \\ S_{IIsph} = -\sum A\bar{A} \cdot y \cdot \Delta(\frac{u}{n}) \\ S_{IIIsph} = -\sum \bar{A}^2 \cdot y \cdot \Delta(\frac{u}{n}) \\ S_{IVsph} = -\sum L^2 \cdot c \cdot \Delta(\frac{u}{n}) \\ S_{Vsph} = -\sum \left\{ \frac{\bar{A}^3}{A} \cdot y \cdot \Delta(\frac{u}{n}) + \frac{\bar{A}}{A} \cdot L^2 \cdot c \cdot \Delta(\frac{1}{n}) \right\} \end{cases}, \tag{3}$$

where $y$ is the height of the marginal ray, $u$ donates the slope of the marginal ray, $n$ represents the refractive index, $c$ is surface curvature, $A$ is the refraction invariant of the marginal ray, $\bar{A}$ is the refraction invariant of chief ray, and $L$ represents the Lagrange invariant of the system. They can be calculated as:

$$\begin{cases} A = n(yc+u) \\ \bar{A} = n(\bar{y}c+\bar{u}) \\ L = n\bar{u}y - nu\bar{y} \end{cases}, \tag{4}$$

where $\bar{y}$ represents the height of the chief ray, and $\bar{u}$ is the slope of the chief ray. The surface with conic coefficient can be expressed by sag [15]:

$$z = \frac{cr^2}{2} + \frac{(\kappa+1)c^3 r^4}{2^2 2!} + \frac{1\cdot 3(\kappa+1)^2 c^5 r^6}{2^3 3!} + \frac{1\cdot 3\cdot 5(\kappa+1)^3 c^7 r^8}{2^4 4!}$$
$$+ \frac{1\cdot 3\cdot 5\cdot 7(\kappa+1)^4 c^9 r^{10}}{2^5 5!} + ... \tag{5}$$

The difference between a conic and a sphere is given by:

$$\Delta z = \frac{[(\kappa+1)-1]c^3 r^4}{2^2 2!} + \frac{1\cdot 3[(\kappa+1)^2-1]c^5 r^6}{2^3 3!} + \frac{1\cdot 3\cdot 5[(\kappa+1)^3-1]c^7 r^8}{2^4 4!}$$
$$+ \frac{1\cdot 3\cdot 5\cdot 7[(\kappa+1)^4-1]c^9 r^{10}}{2^5 5!} + ... \tag{6}$$

In the initial structural design round, we only consider the fourth-order aspheric coefficient $c_4$. In this case, the change of the vector height of the aspheric surface can be expressed as:

$$\Delta z = (\frac{1}{8}\kappa c^3 + c_4)r^4 + ... \tag{7}$$

The terms of the aspheric surface only affect the fourth power of the aperture and the terms of higher order. To the level of third-order aberrations, aspherisation only affects the spherical aberration if the aperture stop is located at the aspheric surface. The contribution to spherical aberration $S_I$ by the aspheric part can be expressed as [16]:

$$\delta S_{Iasph} = \sum 8 \cdot (\frac{1}{8}\kappa c^3 + c_4) y^4 \Delta n. \tag{8}$$

The coordinates of the chief ray on each surface will vary with the location of the pupil. The value of the field curvature $S_{IV}$ is independent of the heights of chief ray and the refraction invariant of the chief ray, so the movement of the aperture does not change the field curvature of the system. The extra coma, astigmatism, and distortion introduced by the aspheric surface when the aperture stop is shifted away from the aspheric surface can be calculated by means of the stop shift formulae:

$$\begin{cases} \delta S_{IIasph} = \frac{\bar{y}}{y} \delta S_{Iasph} \\ \delta S_{IIIasph} = (\frac{\bar{y}}{y})^2 \delta S_{Iasph}, \\ \delta S_{Vasph} = (\frac{\bar{y}}{y})^3 \delta S_{Iasph} \end{cases} \tag{9}$$

where $S_{IIasph}$, $S_{IIIasph}$ and $S_{Vasph}$ indicate the Seidel sums by the aspheric part after a stop shift. Then the Seidel coefficients of the optical system with conic coefficient and fourth-order aspheric coefficient $S_I - S_V$ can be acquired as:

$$\begin{cases} S_I^* = S_{Isph} + \delta S_{Iasph} = S_{Isph} + \delta S_{Iasph1} + \delta S_{Iasph2} \\ S_{II}^* = S_{IIsph} + \delta S_{IIasph} \\ S_{III}^* = S_{IIIsph} + \delta S_{IIIasph} \\ S_{IV}^* = S_{IVsph} \\ S_V^* = S_{Vsph} + \delta S_{Vasph} \end{cases} \tag{10}$$

where the $S_{Iasph1}$ represents the contribution of conic coefficient $\kappa_1$ of PM to spherical aberration, and $S_{Iasph2}$ represents the contribution of conic coefficient $\kappa_2$ and fourth-order aspheric coefficient $c_4$ of SM to spherical aberration. As the aperture is located on the PM, the conic coefficient of the PM only affects the spherical aberration term in the primary aberration.

According to the specifications, $y_1$=110mm, $u_1$=0, $\bar{y}_1 = 0$, and $\bar{u}_1 = 2.1 \times 10^{-4}$. To keep the reasonable space between the surfaces, we set $d_1=d_2=-300$mm. Correspondingly, the corresponding optical distances in paraxial ray tracing are $\tau_1 = d_1/n_2 = 300$ mm and $\tau_2 = d_2/n_3 = 300$ mm. The optical invariant of the system can be calculated as $L$=0.0230 mm. The objective of optimization is to control the primary aberration and the effective focal length of the system, so the evaluation function is set as:

$$F(r_1, r_2, \kappa_1, \kappa_2, c_4) = \sqrt{w_1 S_I^2 + w_2 S_{II}^2 + w_3 S_{III}^2 + w_4 S_{IV}^2 + w_5 S_V^2 + w_6 (f - f_0)^2}, \tag{11}$$

where $w_i$ is the weight coefficients, $f$ is the focal length of the system, and $f_0$ is the target focal length [17]. Then the curvature radii of PM and SM and the distances between them can be derived from the analysis of Seidel coefficients. The detailed parameters are listed in Table 2. The Seidel coefficients are respectively $S_I$=5.32e-6, $S_{II}$=-1.56e-5, $S_{III}$=1.60e-6, $S_{IV}$=1.82e-6, and $S_V$=-4.07e-08.

Table2. Coaxial Initial Structure Parameters of PM and SM

| Surface | Radis (mm) | Distance (mm) | Conic | 4th-order |
|---|---|---|---|---|
| PM(Stop) | -666.668 | -300.000 | -1.000 | |
| SM | -75.001 | 300.000 | -1.404 | -4.689e-08 |
| Image plane | - | - | - | |

*3.2 Design of Tertiary mirror and Quaternary mirror*

As shown in in Fig. 2, we trace the rays in the opposite direction to design the Tertiary mirror (TM) and Quaternary mirror (QM). The pair of mirrors functions as a collimator, and the FOV and the size of the entrance pupil (in fact exit pupil of the whole system) are determined by the magnification of the PM and SM. The height of the marginal ray is scheduled to be 2.75mm. Since the focal points of the two systems overlap, the sum of the back focal distance (BFD) of system consists of PM and SM and the BFD of the system consists of TM and QM systems should be less than the specified 350mm. For TM and QM, the value of BFD is 50mm.

The light ray is traced as a sequence manner, ignoring the occlusion problem, which will be addressed by the operation of off-axis in later designs. The aperture and the FOV of this sub-system are small. Therefore, the SM is configured as a sphere, and the QM is set as a conical surface. The primary aberrations are also calculated by Eqs. (3)-(10). The detailed parameters are listed in Table 2. The Seidel coefficients are respectively $S_I$=2.38e-05, $S_{II}$=1.17e-05, $S_{III}$=7.33e-06, $S_{IV}$=8.98e-06, and $S_V$=-6.86e-07.

Table3. Coaxial Initial Structure Parameters of TM and QM

| Surface | Radis (mm) | Distance (mm) | Conic |
|---|---|---|---|
| Stop | - | 200 | |
| PM | -690.084 | -114.983 | -10.000 |
| SM | 127.791 | 50.000 | 0 |
| Image plane | - | - | - |

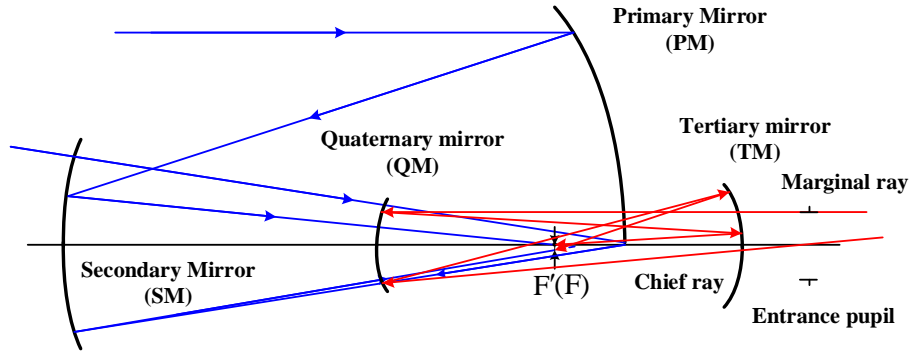

Fig. 2. Ray tracing in initial coaxial system of TM and QM

## 4. Design and Optimization

Firstly, the coaxial afocal system is transformed to an off-axis system with conic surfaces [18]. The imaging quality of Tianqin telescope is evaluated by RMS wavefront error during the optimization process, which is different from a conventional imaging telescope. In addition, the wavefront error at the sampled FOVs in all directions should be kept as low as possible. The RMS wavefront error is often taken as the performance metric to evaluate the variation of an actual wavefront map from a perfect spherical wavefront, and is defined as: [19]:

$$W_{rms} = \sqrt{\frac{1}{A} \iint [W(x,y) - W_{mean}(x,y)]^2 \, dxdy}, \quad (12)$$

where $A = \iint dxdy$, and $W_{mean}$ is the average wavefront deviation over the pupil area, which is expressed as:

$$W_{mean}(x,y) = \frac{1}{A} \iint W(x,y) \, dxdy. \quad (13)$$

Preliminary analysis indicates that the straightforward optical design is very difficult to build in an optical design software. Therefore, a progressive strategy is adopted during our design process, and the number of the aspheric terms gradually increases.

In the first step of the optimization, proper constraints should be selected and applied [7,19-20]. The axial optical length between the SM and TM is constrained up to 370 mm to satisfy the fabrication requirement. Take the layout plan of the refraction-type optical system into consideration, the distance between the primary and secondary mirrors is limited to less than 300 mm. At the same time, in order to control the weight of the supporting structure of the system, the off-axis amount in the y-direction should not be too large while eliminating the mirror occlusion. To accommodate the interface to the optical bench, the coordinates of the rays striking the image plane are constrained to control the incidence of light from the exit pupil being approximately perpendicular to the optical bench.

In the following steps, the commercial optical design software is employed for the further optimization. PM surface is set as the global coordinate reference surface. An off-axis configuration in the y direction is set to eliminate the occlusion based on the coaxial parameters of two Cassegrain systems obtained in the previous section. By controlling the combined system power $\phi_{M1M2}$ and $\phi_{M3M4}$, the reduction ratio of laser beam can be determined as $m = \phi_{M3M4} / \phi_{M1M2}$. Only an initial structure with potential has been constructed here, as the ability to correct off-axis aberrations by configurations with only conic surfaces is limited. The parameters are listed in Table 4, including the curvature radii, distance, conic coefficient, and the decenter value. The primary mirror is about F/1.5, and the parameters of the primary mirror will not be optimized in the subsequent design process. The RMS wavefront error of the marginal FOV is 0.00162 λ.

Table 4. Off-axis initial structure parameters

| Surface | Radis (mm) | Distance (mm) | Conic | Decenter Y(mm) |
|---|---|---|---|---|
| PM(Stop) | -666.661 | -300.000 | -1.000 | -130 |
| SM | -74.994 | 352.711 | -1.561 | - |
| TM | -150.357 | -74.592 | 3.673 | -5 |
| QM | 492.640 | - | -8.641 | -25 |

Then we first introduce even aspheric surface on SM. Higher-order aspheric coefficient are gradually set to be the optimized parameters for further optimization. In addition, the weight coefficients of the marginal FOVs should be properly improved to make the aberration evenly distribute across the full FOV. Because the tilt-to-length (TTL) noise generated by the non-uniform wavefront error cannot be eliminated by the calibration method, and it will affect the measurement results inevitably. At this stage, TM and QM have no tilt, as the tilt will introduce additional asymmetric aberrations. After optimization, the aberrations induced by the decenter are partly corrected. The basic design parameters and aspheric coefficients of M2 are listed in Table 5.

Table 5. Structure parameters of the design with M2 asphere

| Surface | Radis (mm) | Distance (mm) | Conic | Asphere coefficient of M2 | |
|---|---|---|---|---|---|
| PM(Stop) | -666.661 | -300.000 | -1.000 | Item | Coefficient |
| SM | -74.994 | 356.374 | -0.736 | 4th order term | 2.443e-07 |
| TM | -155.3009 | -68.091 | 4.320 | 6th order term | -5.868e-12 |
| QM | 540.424 | - | -8.669 | 8th order term | 3.451e-17 |

In the subsequent stage, optimization is used surface by surface to minimize the design residual aberrations at each surface. The parameters including the radius, the aspherical coefficients of the TM and QM, decenter in Y direction, tilt about X axis are set as variables in the optimization. When the configuration becomes stable, the additional constraint about the exit pupil should be considered, where the exit pupil must be positioned between the telescope and the optical platform, the same size as the local laser beam. According to the characteristics of the afocal optical system, the exit pupil is controlled by ray tracing from the central FOV. With the optimization being conducted, the location and size of the exit pupil are restricted to the specification values, and the higher order terms of aspheric coefficients are gradually added to improve the imaging quality.

The whole design procedure adopts two optimized strategies, as illustrated in Fig. 3. It is clear that the large off-axis aberrations introduced by the off-axis mirrors are difficult to control with the traditional conic surfaces. Therefore, it is necessary to introduce higher order aspherical coefficients into these surfaces. However, only two aspheric surfaces with 4th-order, 6th-order, and 8th-order coefficients cannot entirely compensate for these aberrations, no matter they are introduced into the SM and TM, or into the SM and QM. The results suggest that the final design is inevitable to use three aspherical mirrors simultaneously. After the number of used aspherical surfaces and the order of each aspherical coefficient are determined, these coefficients are set as variables at the same time for final optimization in the software, and the final design result is obtained.

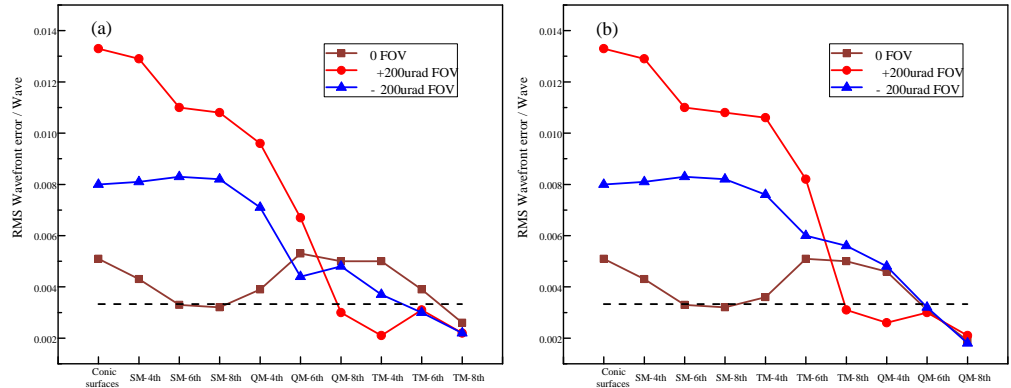

Fig. 3. The curve of RMS wavefront error with each aspheric coefficient at the FOV of 0 and ±200μrad: (a) SM-QM-TM (b) SM-TM-QM

## 5. Design results and performance

The final design system is illustrated in Fig. 4, which consists of a large diameter paraboloid mirror and three aspheric surface mirrors. The structural layout of the system is adjusted reasonably, where occlusion is completely eliminated. The basic structural parameters and even-order aspheric coefficients are detailed in Table 6 and Table 7. The total optical length is 328.151mm, and the transverse length is 260.044mm in the Y direction. The exit pupil is located 100 mm behind the QM along the z-axis. An intermediate image plane is arranged between SM and TM so that the field stop can be placed to shield large amounts of stray light.

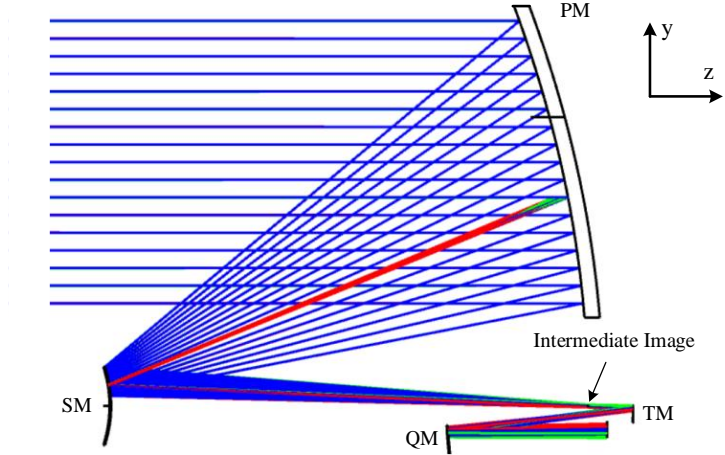

Fig. 4. Optical layout of the recommended Tianqin telescope design

Table 6. Off-axis structure parameters of recommended design

| Surface | Radis (mm) | Distance (mm) | Conic | Decenter Y(mm) | Tilt (°) |
| --- | --- | --- | --- | --- | --- |
| PM(Stop) | -666.661 | -300.000 | -1.000 | -130 | - |
| SM | -74.994 | 328.155 | 0.032 | - | - |
| TM | -94.556 | -114.983 | -1.893 | -3.998 | - |
| QM | 366.344 | - | -7.904 | -24.485 | -5.306 |

Table 7. Detailed even-order aspheric coefficients of recommended design

| Surface | 4th-order term | 6th-order term | 8th-order term |
| --- | --- | --- | --- |
| SM | 4.714e-07 | 2.139e-11 | 2.959e-15 |
| TM | -2.446e-06 | 3.491e-08 | - |
| QM | -6.765e-08 | 1.755e-10 | -1.431e-13 |

In order to evaluate the image quality of an afocal system, we insert a perfect lens at the pupil with an ideal lens to focus the collimated rays. The standard spot diagrams on the image plane of all configurations are exhibited in Fig. 5. The figure shows that the RMS radii of the spots at all FOV are much smaller than that of the Airy spot, and the principal components of residual aberrations are spherical aberration and coma. Fig. 6 demonstrates the wavefront aberration at the full FOV. The RMS wavefront error of the system is not greater than $\lambda/400$, ensuring that the laser beam after beam expansion has high quality of wavefront. The image distortion is illustrated in Fig. 7. The magnitude of the maximum distortion is about -0.06% at the margin fields.

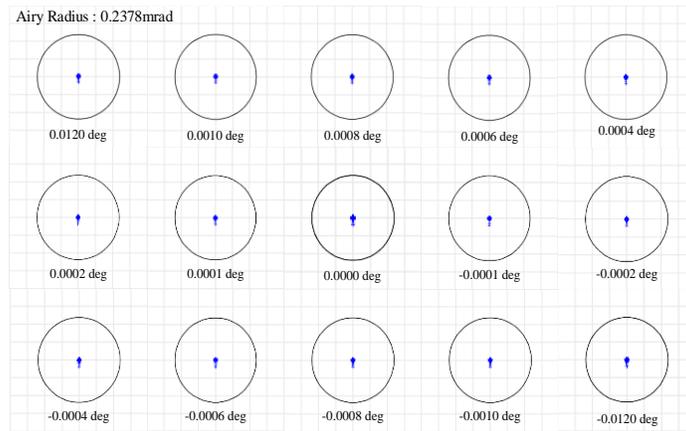

Fig. 5. The spot diagram of the recommended Tianqin telescope design

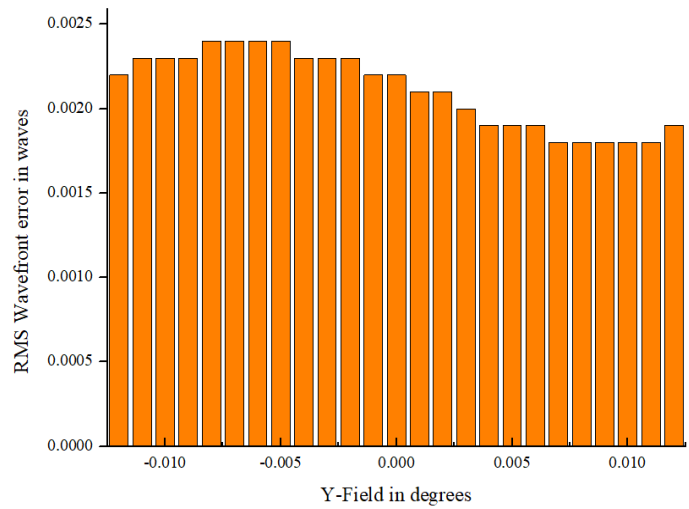

Fig. 6. RMS wavefront error vs Y-field

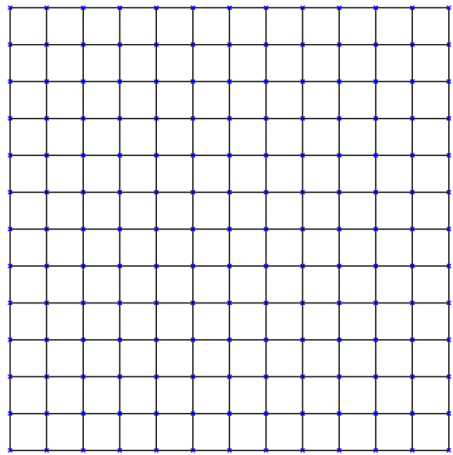

Fig. 7. Distortion of the recommended Tianqin telescope design

## 6. Sensitivity analysis and stray light analysis

Considering the existence of aberrations in the optical elements on the optical bench, leaving a certain margin for the assembly of the interferometer optical system, the overall wavefront error goals of the telescope system are $\lambda/30$. In addition to the six flight telescopes aboard the three spacecrafts, prototypes need to be built for testing on the ground. Fabrication of these telescopes on a small scale requires verification of the telescope's tolerance performance to minimize schedule risk and engineering costs. Using the optical design software's wavefront tolerance analysis function, the sensitivity of the optical system to manufacturing and assembly errors can be evaluated.

The wavefront error of the telescope is dominated by the surface figure error of the mirrors and the assembly alignment. The surface error is caused by machining error as well as the deformation due to the assembly. After a rough sensitivity analysis, we know that the wavefront error sensitivities of the TM and QM positions in this design are low. They are actually considered to be mounted on a translation stage, which can at least partially compensate the defocus of the system. Therefore, the positions of TM and QM serve as the compensator in sensitivity analysis to obtain the maximized performance at final assembly process. The compensation range is set to $\pm 3mm$. The tolerance distribution of each term is listed in Table 8.

**Table 8. Tolerance Setting in Sensitivity analysis**

| Surface | Tolerances | | | | | |
|---|---|---|---|---|---|---|
| | Radius | Thickness | Decenter in x | Decenter in y | Tilt in x | Tilt in y |
| PM | $\pm 5\mu m$ | $\pm 10\mu m$ | $\pm 0.5\mu m$ | $\pm 0.5\mu m$ | 0.06arc min | 0.06arc min |
| SM | $\pm 10\mu m$ | $\pm 20\mu m$ | $\pm 0.5\mu m$ | $\pm 0.5\mu m$ | 0.09arc min | 0.09arc min |
| TM | $\pm 10\mu m$ | $\pm 20\mu m$ | $\pm 10\mu m$ | $\pm 10\mu m$ | 0.6arc min | 0.6arc min |
| QM | $\pm 10\mu m$ | $\pm 20\mu m$ | $\pm 10\mu m$ | $\pm 10\mu m$ | 0.6arc min | 0.6arc min |

A Monte Carlo tolerance analysis is executed to predict the optical performance. Each parameter is perturbed with a random value within the tolerance range for each variable. This procedure is repeated for 1000 times, and each result for the RMS wavefront error are derived in the sensitivity analysis. The overall analysis results are listed in Table 9. The worst-performing result is $0.0304\lambda$, which meets the wavefront error requirements of the whole system. There is an 98% probability that the RMS wavefront error is less than $0.02419 \lambda$, which is acceptable and leaves room for other possible error source.

**Table 9. Sensitivity Analysis Results**

| | Worst | 98% | 90% | 80% | 50% | 20% | 10% | 2% |
|---|---|---|---|---|---|---|---|---|
| RMS WFE | 0.03038 | 0.02419 | 0.02121 | 0.01783 | 0.01201 | 0.00651 | 0.00425 | 0.00267 |

Super-low stray light requirement is mainly set by two factors. One is that the signal power received by the telescope is about $10^{-10}$ of the transmitted, while the back-scattering light is produced by local high power laser beam from optical bench. Another is that coherent detection scheme is implemented by creating an interference signal with small incoming signal and higher power local reference laser beam, but it also implies that the detector is very sensitive to stray light. Here we only analyze the scattered light from inside the optical system, caused by the surface roughness and the particulates of the surface [21-22]. The scattering model uses Harvey Shake model, and the simulation is completed in commercially software ASAP. The scatter model is applied to all of the mirror surfaces, and these surfaces are all are coated with a 0.99 reflectivity film. In the simulation model established, there is no other baffle except the field stop placed in front of the detector. Then the stray light received by the detector in the field of view of the telescope is simulated numerically. Cleanliness levels of CL200 and surface roughness of 0.5 nm can meet the design requirements. The design and optimization of the field stop and stray light shield will be explored in future work.

## 7. Conclusion

In this paper, we have successfully designed a high performance space-borne telescope with ultra-low wavefront distortion, super-low stray light level as well as ultra-stable structure for space-based interferometric gravitational-wave detectors, such as Tianqin project. Our provided system design can enjoy an advantage of compactness as well as high efficiency with no occlusion for its full FOV. During the design process, we have conducted detailed discussion with regards to the initial structure design, structure constrains as well as the optimization strategy. In addition, comprehensive performance analysis has been performed to verify our provided design, such as wavefront imaging quality, the exit pupil location, sensitivity analysis, and stray light suppression level, and so on. Our analysis demonstrates that the design can have relatively high performance with the RMS wavefront errors of full FOV less than $\lambda/300$ as well as super low-level back reflection under normal optical processing and cleaning level, which satisfies TianQin's science requirements very well. Our performed tolerance analysis results for the designed telescope also indicate that the telescope can have very good manufacturability. In conclusion, our designed space-borne telescope can satisfy the strict requirements of the space-borne gravitational-wave detection mission very well due to its relatively high performance in almost all respects and our provided design strategy can also be a good guidance for any other similar science project like LISA and TaiJi.

### Disclosures

The authors declare no conflicts of interest.